\documentstyle [aps,psfig]{revtex}

\begin{document}
\draft
\title{Numerical Evidence that the Singularity in Polarized $U(1)$
Symmetric Cosmologies on $T^3 \times R$ is Velocity Dominated \thanks{
e-mail: berger@oakland.edu, moncrief@hepvms.physics.yale.edu}} 
\author{Beverly K. Berger}
\address{Department of Physics, Oakland University, Rochester, MI 48309 USA}
\author{Vincent Moncrief}
\address{Departments of Physics and Mathematics, Yale University, New Haven, CT
06520 USA}

\maketitle
\bigskip
\begin{abstract}
Numerical evidence supports the conjecture that polarized $U(1)$ symmetric
cosmologies have asymptotically velocity term dominated singularities.

\end{abstract}
\pacs{98.80.Dr, 04.20.J}

\section{Introduction}
The cosmological singularity in spatially homogeneous models is known to be
either asymptotically velocity term dominated (AVTD) (Kasner-like)
\cite{eardley72,isenberg90} or Mixmaster-like \cite{belinskii71b,misner69}. 
Analytic \cite{isenberg90,grubisic93} and numerical studies
\cite{berger93,berger97b} have shown that the singularity in the spatially
inhomogeneous Gowdy models
\cite{gowdy71} is AVTD everywhere in polarized models and everywhere except,
perhaps, at a set of measure zero in the generic, unpolarized case. Long
ago, Belinskii, Khalatnikov, and Lifshitz (BKL) \cite{belinskii71b} argued that the
generic singularity in Einstein's equations is locally of the Mixmaster type. This
remains controversial \cite{barrow79} but serves as a conjecture which can be
tested. The Gowdy models, though interesting in their own right, have two commuting
Killing fields which precludes local Mixmaster behavior for $T^3 \times R$
topology. (For generalized Gowdy models which appear to exhibit
Mixmaster dynamics see
\cite{weaver98}.) However, the more general one Killing field
$U(1)$ symmetric cosmologies on $T^3 \times R$ should generically have
Mixmaster-like singularities if the BKL conjecture is correct.

An AVTD solution to Einstein's equations is obtained by neglecting all terms
containing spatial derivatives. The parameters of the solution to the resultant
ODE's are then assumed to depend on the spatial coordinates. An AVTD
singularity is then one where the solution to the full Einstein equations comes
arbitrarily close to an AVTD solution as the singularity is approached
\cite{isenberg90}. For the $U(1)$ cosmologies, the AVTD equations may be solved
exactly. However, one may worry about the coordinate and slicing dependence of the
notions of space and time leading to such dependence in our definition of AVTD. We
shall argue here that the AVTD behavior of the singularity is not strongly slicing
dependent.

We have already shown \cite{berger93}
that numerical methods used in the Gowdy cosmologies can be easily generalized
to the $U(1)$ case. However, numerical difficulties associated with spatial
differencing in two dimensions have prevented so far a complete understanding
of generic $U(1)$ models probably because Gowdy-like spiky features \cite{berger97b}
cannot yet be modeled with sufficient accuracy. However, the subclass of polarized
$U(1)$ models does not appear to develop spiky features and thus can be treated
well by our numerical methods. We consider vacuum $U(1)$ symmetric cosmologies
on $T^3 \times R$. These can be described by two variables $\varphi$ and
$\omega$ analogous to the Gowdy wave amplitudes for the $+$ and $\times$
polarizations of gravitational waves and three variables $\Lambda$, $x$, and
$z$ which contain nondynamical information. Polarized $U(1)$ models have the
degree of freedom associated with $\omega$ set equal to zero. This condition is
preserved for all time both by Einstein's equations and by the discrete form
thereof used in our numerical simulations. While a general solution to the initial
value problem is not known explicitly, we have found an algebraic solution to the
constraints which contains three (four for generic $U(1)$ models) arbitrary
functions of the spatial variables and is sufficiently general to yield generic
evolution toward the singularity within the polarized $U(1)$ class of models. We
find that the evolution toward the singularity is AVTD everywhere as had previously
been conjectured
\cite{grubisic94}.  This conclusion is based (a) on the exponential decay with time
of all terms in the Hamiltonian containing spatial derivatives, (b) the exponential
decay with time of the change in certain functions of the spatial coordinates which
are constant in time in the AVTD regime, and (c) fits of the behavior of the
variables at randomly selected spatial points to the expected linear or constant
asymptotic time dependence. 

\section{The Model}
The spacelike Killing vector $\xi = \partial / \partial x^3$ of the $U(1)$
symmetry allows a generic metric in this class to be described in terms of the
coordinate $x^3$ along
$\xi$ and coordinates $u$ and $v$ and $\tau$ parametrizing the quotient $2 + 1$
manifold
\cite{moncrief86}. Using notation from \cite{berger93} and taking $u\in [0,2\pi ]$
and $v
\in [0,2
\pi]$ the $3 + 1$ metric is
\begin{equation}
\label{metric}
ds^2=e^{-2\varphi }\left\{ {-N^2e^{-4\tau }d\tau ^2+e^{-2\tau }e^\Lambda
e_{ab}dx^adx^b} \right\}+e^{2\varphi }\left( {dx^3+\beta _adx^a} \right)^2
\end{equation}
where $\{x^a\} = \{u,v\} = \{x^1,x^2\}$, the lapse $N = e^\Lambda$, and $g_{ab} =
e^\Lambda e_{ab}$ where
\begin{equation}
\label{eab}
e_{ab}={1 \over 2}\left[
{\matrix{{e^{2z}+e^{-2z}(1+x)^2}&{e^{2z}+e^{-2z}(x^2-1)}\cr
{e^{2z}+e^{-2z}(x^2-1)}&{e^{2z}+e^{-2z}(1-x)^2}\cr }} \right]
\end{equation}
is the conformal metric of the $u$-$v$ plane. 

In the polarized case, $\beta_a = 0$ but for
the general case we must construct $\{ \beta_a \} = \{ \beta_1,\beta_2 \}$ as
follows: At an initial time, $\tau = \tau_0$, define
\begin{eqnarray}
\label{beta1}
\beta _1^*(u,v)&=&c_{1}+\int_0^v {dv'\left\{ {r(u,v')-{1 \over {2\pi
}}\int_0^{2\pi } {dv''[r(u,v'')]}} \right\}}, \nonumber\\
  \beta _2^*(u,v)&=&c_2-{1 \over {2\pi }}\int_0^u {du'\int_0^{2\pi }
{dv'\,r(u',v')}}
\end{eqnarray}
where $c_1$ and $c_2$ are constants. These constants are like ``twist''
constants \cite{moncrief86} and are arbitrary (but physically significant). These
$\beta^*$'s will be periodic on $T^2$ provided
\begin{equation}
\label{rcondition}
\int_0^{2\pi} \,du \,\int_0^{2\pi}\, dv \,\, r(u,v) = 0.
\end{equation}
Now let $\beta_1 = \beta_1^* + {\partial \lambda }/ {\partial u}$ and 
$\beta_2 = \beta_2^* + {\partial \lambda }/ {\partial v}$ where $\lambda$ is
arbitrary and could be identically zero to give $\{\beta_a\}$ at $\tau_0$. Finally,
define $\beta_a(u,v,\tau)$ by computing
\begin{equation}
\label{definebeta}
\beta_a(u,v,\tau) = \beta_a(u,v,\tau_0) - \int_{\tau_0}^\tau \,d\tau'
\,\,e^{-2\tau'} N e^{-4 \varphi} e_{ab} \varepsilon^{bc} \omega,_c
\end{equation}
where
\begin{equation}
\label{antisym}
\varepsilon ^{bc}=\left( {\matrix{0&1\cr
{-1}&0\cr
}} \right).
\end{equation}
As in \cite{moncrief86}, a canonical transformation replaces the terms in the
Einstein-Hilbert action $e^a \, \dot \beta_a + \beta_0 \, e^a,_a$ by $r \, \dot
\omega$ where the momentum $e^a$ conjugate to $\beta_a$ identically
satisfies the constraint $e^a,_a = 0$ if $e^a = \varepsilon^{ab}\,\omega,_b$.
Einstein's equations may be obtained by the variation of
\begin{eqnarray}
\label{Hu1}
H &=& \int \int du \kern 1pt dv \,{\cal H} \nonumber \\
&=& \int \int du \kern 1pt dv \left( {1 \over 8}p_z^2+{1 \over 2}
e^{4z}p_x^2+{1 \over 8}p^2+{1 \over 2}e^{4\varphi }r^2-{1 \over 2}p_\Lambda
^2+2p_\Lambda  \right) \nonumber \\
&& +e^{-2\tau } \int \int du \kern 1pt 
dv \left\{  \left( {e^\Lambda e^{ab}} \right) ,_{ab}- \left( {e^\Lambda e^{ab}}
\right) ,_a\Lambda ,_b+e^\Lambda  \right. \left[  \left( {e^{-2z}}
\right) ,_u x,_v- \left( {e^{-2z}} \right) ,_v x,_u \right] \nonumber \\
&& \left. +2e^\Lambda e^{ab}\varphi ,_a\varphi ,_b+{1 \over 2}
e^\Lambda e^{-4\varphi }e^{ab}\omega ,_a\omega ,_b \right\} \nonumber \\
&=& H_K+H_V = \int \int du \kern 1pt dv \, {\cal H}_K +\int \int du \kern 1pt dv
\,V.
\end{eqnarray}
The Hamiltonian and momentum constraints are respectively
\begin{equation}
\label{H0}
{\cal H}^0 = {\cal H} - 2 p_\Lambda = 0
\end{equation}
and
\begin{eqnarray}
\label{Hu}
{\cal H}^u&=&p_z\,z,_u+p_x\,x,_u+p_\Lambda \,\Lambda ,_u-p_\Lambda ,_u+p\varphi
,_u+r\omega ,_u+{1 \over 2}\left\{ {\left[ {e^{4z}-(1+x)^2}
\right]p_x-(1+x)p_z} \right\},_v \nonumber \\
  & &-{1 \over 2}\left\{ {\left[ {e^{4z}+(1-x^2)} \right]p_x-x\kern 1pt p_z}
\right\},_u=0,
\end{eqnarray}
\begin{eqnarray}
\label{Hv}
{\cal H}^v&=&p_z\,z,_v+p_x\,x,_v+p_\Lambda \,\Lambda ,_v-p_\Lambda ,_v+p\varphi
,_v+r\omega ,_v-{1 \over 2}\left\{ {\left[ {e^{4z}-(1-x)^2}
\right]p_x+(1-x)p_z} \right\},_u \nonumber \\
 & &+{1 \over 2}\left\{ {\left[ {e^{4z}+(1-x^2)} \right]p_x-x\kern 1pt p_z}
\right\},_v=0.
\end{eqnarray}
To evolve Einstein's equations numerically, we need to solve the initial value
problem. While an explicit general solution is not available, a particular class of
solutions may be obtained as follows: To solve the momentum constraints
(\ref{Hu}) and (\ref{Hv}) set $p_x = p_z = \varphi,_a = \omega,_a = 0$ to leave
$p_\Lambda \, \Lambda,_a - p_\Lambda,_a = 0$ which may be satisfied by
requiring $p_\Lambda = c e^\Lambda$. For sufficiently large $c$, the
Hamiltonian constraint may be solved algebraically for either $p$ or $r$. In
general, this leaves as free data the four functions $x$, $z$, $\Lambda$, and
either $r$ or $p$.  However, in the
polarized case, we demand
\begin{equation}
\label{polcond}
\omega = r = 0
\end{equation}
so that the Hamiltonian constraint must be solved for $p$ with three arbitrary
functions $x$, $z$, and $\Lambda$. We note that while any $c$ will solve the
initial value problem, to approach the singularity we require $c > 0$ in order to
give $p_\Lambda > 0$ so that $\Lambda$ will decrease (become more negative) as
$\tau \to \infty$. This is required because the determinant of the metric has a
factor $e^{2\Lambda}$ which measures the area in the $u$-$v$ plane.

The AVTD equations may be obtained by variation of $H_K$ in (\ref{Hu1}) and
have the exact solution (see \cite{berger93,berger97a})
\begin{eqnarray}
\label{avtdsoln}
z&=&-v_z(\tau -\tau _{0z})+\ln [|\mu_z| \,(1+e^{-4v_z(\tau -\tau _{0z})})] \to
-v_z\tau
\quad {\rm as}\;\tau \to \infty , \nonumber \\
x&=&\xi_z -\left[ {\mu_z \,(1+e^{-4v_z(\tau -\tau _{0z})})} \right]^{-1}\quad \quad
\quad
\;\to x_0\;\quad {\rm as}\;\tau \to \infty , \nonumber \\
p_z&=&-4v_z{{\,(1-e^{-4v_z(\tau -\tau _{0z})})} \over {\,(1+e^{-4v_z(\tau -\tau
_{0z})})}}\quad \quad \quad \quad \quad \quad \to -4v_z\quad \;\;{\rm as}\;\tau \to
\infty, \nonumber  \\
p_x&=&-\mu_z \,v_z \equiv p_x^0, \nonumber  \\
\varphi&=&-v_\varphi(\tau -\tau _{0\varphi})+\ln [|\mu_\varphi|
\,(1+e^{-4v_\varphi(\tau -\tau _{0\varphi})})]
\to -v_\varphi\tau
\quad {\rm as}\;\tau \to \infty , \nonumber  \\
\omega&=&\xi_\varphi -\left[ {\mu_\varphi \,(1+e^{-4v_\varphi(\tau -\tau
_{0\varphi})})}
\right]^{-1}\quad \quad \quad
\;\to \omega_0\;\quad {\rm as}\;\tau \to \infty , \nonumber \\
p&=&-4v_\varphi{{\,(1-e^{-4v_\varphi(\tau -\tau _{0\varphi})})} \over
{\,(1+e^{-4v_\varphi(\tau -\tau _{0\varphi})})}}\quad \quad \quad \quad \quad \quad
\to -4v_\varphi\quad \;\;{\rm as}\;\tau \to
\infty,  \nonumber  \\
r&=&-\mu_\varphi \,v_\varphi \equiv r^0, \nonumber  \\
\Lambda &=& \Lambda_0 +(2 - p_\Lambda^0)\tau,  \nonumber  \\
p_\Lambda &=& p_\Lambda^0
\end{eqnarray}
subject to the AVTD limit of the Hamiltonian constraint (\ref{H0})
\begin{equation}
p_\Lambda^2 = {1 \over 4} p_z^2 + e^{4z} p_x^2 + {1 \over 4} p^2 + e^{4\varphi}
r^2.
\end{equation}
The AVTD limit is expressed in terms of $\mu_z$, $v_z \ge 0$, $\xi_z$,
$\tau_{0z}$, $\mu_\varphi$, $v_\varphi \ge 0$, $\xi_\varphi$, and
$\tau_{0\varphi}$ which are functions of $u$ and $v$. The limits on $v_z$ and
$v_\varphi$ arise in order for the limiting forms of (\ref{avtdsoln}) to cause the
exponents in (\ref{Hu1}) to decay as $\tau \to \infty$. (See the discussion of
similar terms in Gowdy models \cite{grubisic93,berger97b}.) The AVTD solution
(\ref{avtdsoln}) may be inverted to give these parameters in terms of the original
variables. Useful examples are
\begin{equation}
\label{vdefine}
v_z = \sqrt{{1 \over 16} p_z^2 + {1 \over 4} e^{4z} p_x^2} \quad , \quad v_\varphi =
\sqrt{{1 \over 16} p^2 + {1 \over 4} e^{4\varphi} r^2} \ \ .
\end{equation}
We further note
\cite{grubisic93} that, in addition to $v_z$, $v_\varphi$, $p_x$, $r$, and
$p_\Lambda$,
\begin{equation}
c_z = {1 \over 2} p_z + p_x x  \quad , \quad c_\varphi = {1 \over 2} p + r \omega
\end{equation}
are also constant in time in an AVTD regime. 

The method of Grubi\u{s}i\'{c} and Moncrief (GM) \cite{grubisic93} can be followed to
determine if one expects the singularity to be AVTD in the polarized model. We
require every term in $V$ (from (\ref{Hu1})) to decay exponentially as $\tau \to
\infty$ if we substitute the limiting AVTD solution, (\ref{avtdsoln}), for the
$U(1)$ variables. Clearly, the behavior will depend on the behavior of the
exponentials. First, as in the Gowdy case \cite{grubisic93,berger97b}, we expect the
remaining nonlinear term in
${\cal H}_K$,
$e^{4z}p_x^2/2$ to drive $p_z$ to negative values yielding $z$ eventually large and
negative. In that case, terms in
$e^{ab}$ proportional to $e^{-2z}$ will dominate. From (\ref{Hu1}), we must examine
the factor $e^{-2 \tau + \Lambda - 2 z}$ which will decay as $\tau \to \infty$ if
$ p_\Lambda - 2 v_z > 0$. (Recall that $v_z \ge 0$.) But the AVTD limiting form of
the Hamiltonian constraint requires
\begin{equation}
\label{avtdh0}
p_z^2 + p^2 -4 p_\Lambda^2 = 0 = 16 v_z^2 + p^2 -4 p_\Lambda^2
\end{equation}
which, in turn, requires $p_\Lambda \ge 2 v_z$. Thus, $p_\Lambda > 0$ is
expected if the singularity is AVTD and yields consistent behavior.

\section{Numerical Methods}
To integrate Einstein's equations, the symplectic method described in detail in
\cite{berger93} is applied to $U(1)$ symmetric cosmologies. For a system described
by a Hamiltonian
\begin{equation}
\label{hsum}
H = H_1 + H_2,
\end{equation}
the corresponding evolution operator $U(\Delta \tau)$ which evolves data
from $\tau$ to $\tau + \Delta \tau$ generated by $H$ can be written as
\cite{fleck76,moncrief83}
\begin{equation}
\label{evolop}
U(\Delta \tau) = U_1({{\Delta \tau} \over 2}) U_2(\Delta \tau) U_1({{\Delta \tau}
\over 2}) +{\cal O}[ (\Delta \tau)^3]
\end{equation}
where $U_1,_2$ are the evolution operators generated by $H_1,_2$. Suzuki
\cite{suzuki90,suzuki91} has shown how to generalize this algorithm to an arbitrary
order in
$\Delta \tau$. This method is useful if the sets of equations of motion obtained by
the separate variations of $H_1$ and $H_2$ are exactly solvable. 

We see \cite{berger93} that (\ref{Hu1}) for the $U(1)$ models has exactly solvable
equations for $H_K$ and $H_V$. We have already obtained the AVTD solution
(\ref{avtdsoln}) from $H_K$'s equations of motion. Since $H_V$ contains no momenta,
the configuration variables $x$, $z$, $\Lambda$, $\varphi$, and $\omega$ are
constants of the motion so the equations are trivially solved. Non-trivial,
however, is the representation of the gradients of $V$ which arise during the
variation. Accurate representation of spatial derivatives in two spatial dimensions
is more difficult than in one dimension since the function of interest is no longer
guaranteed to vary along the differencing direction. These inaccuracies have been
found to limit the duration of {\it generic} $U(1)$ model simulations {\it but are
not problematical for polarized models}. This is almost certainly due to the fact
that Gowdy-like, steepening spiky features characterized by increasingly (with
$\tau$) large spatial gradients are absent in polarized $U(1)$ models as they also
are in polarized Gowdy models. We emphasize that the ``averaging'' mentioned in
\cite{berger97a} as necessary to stabilize generic $U(1)$ evolution is not required
in the polarized case.

To illustrate the behavior of typical polarized $U(1)$ symmetric models, we consider a
particular choice of initial data with
$\Lambda = A \sin u \sin v$, $x = z = \cos u \cos v$, and $p_\Lambda = c e^\Lambda$
with $A = 1$, $c = 14$. Other choices of initial data yield qualitatively
similar results.
 
The constraints are imposed initially and monitored thereafter.
Fig.~1 shows the maximum value of the constraints on the spatial grid vs time and
spatial resolution.Although it is not clear how small
one should require the constraints to be, the demonstrated convergence allows one to
argue that the small constraint violation can be neglected in our interpretation of
the results.  Choptuik has argued that constraint convergence is an indicator
of convergence of the numerical solution to the true solution to
Einstein's equations \cite{choptuik91}. 

\section{Results}
Fig.~2 displays the maximum value of $\log_{10} V$ from (\ref{Hu1}) at two spatial
points as a function of $\tau$. One expects this quantity to vanish as $\tau \to
\infty$ for an AVTD singularity. Here we demonstrate that it vanishes exponentially
as one expects. We note that the exponential decay of $V$ measures only
consistency and does not prove AVTD behavior. The observed exponential decay is
compared to the predicted $e^{(-p_\Lambda+p_z/2)\tau}$ behavior. While only
representative spatial points have been shown, exponential decay at the predicted
rate is seen at all spatial points. 

In Fig.~3, the
maximum values over the spatial grid of the {\it change} with time in the AVTD
functions that should be constant in time in the AVTD limit---$p_x$, $v_z$,
$v_\varphi$, $c_z$, $c_\varphi$,
$p_\Lambda$--- are plotted vs $\tau$. Again, one sees exponential decay to the
level of machine noise. (This machine noise does not show up in $\log_{10} V$ which
is an evaluation but does appear when differences are computed.) These quantities
are strictly constant in the AVTD limit. In order to compare the observed to
predicted decay of these quantities, one needs to go beyond the AVTD solution. In
the Gowdy case, this was done by GM \cite{grubisic93}. 

Finally, in Fig.~4, the variables $\Lambda$, $z$, and
$\phi$, expected to be linear with $\tau$ in the AVTD limit, and $x$ expected to be
constant in
$\tau$ are shown vs $\tau$ at typical spatial points. For the first three, linear
fits are shown. We see that $x$ does not show the constant in $\tau$ behavior one
would expect as $\tau \to \infty$. However, we note from Fig.~4 that $x \approx
0$. This small constant value allows the exponentially decaying term in $x$ to be
measurable. The lines in Fig.~4(d) are fit by a constant plus an 
exponential decaying with the expected rate of $4 v_z$ from
(\ref{avtdsoln}). Again, although only representative points are displayed, the same
behavior is seen at all spatial points.

\section{Discussion}
Numerical studies of polarized $U(1)$ models provide strong evidence in support of
their conjectured AVTD singularity. For a representative choice of initial data,
all relevant terms are examined for consistency with AVTD behavior. Terms predicted
to decay exponentially in the AVTD limit as $\tau \to \infty$ do so with the
expected slope. The terms $c_z$, $c_\varphi$, $v_z$, $v_\varphi$, $p_x$, and
$p_\Lambda$ are strictly constant in $\tau$ for all $\tau$ in the AVTD solution.
Decay to the observed constant values could be compared with the prediction from
the next order in an asymptotic expansion where the AVTD limit is the zero order
term \cite{grubisic93}. This has not been done. However, this is not required to
demonstrate that the singularity is AVTD.

Given the consistency of our numerical results with the AVTD limit, we argue that
longer duration simulations and other initial data sets will reveal nothing new.
This is because the method of GM suggests that exponential decay of all the terms
in $V$ is consistent with the AVTD limit as $\tau \to \infty$. As shown in
the Gowdy case \cite{grubisic93} and, of course, in Mixmaster itself
\cite{belinskii71b}, the exponential growth of such terms is required to change the
qualitative behavior of the model at a given spatial point. In the absence of this
growth, there is no mechanism within Einstein's equations that could destroy the
AVTD behavior.

Finally, we consider the possible slicing dependence of the notion of AVTD. First,
we note that the canonical transformation to $r$ and $\omega$ has interchanged
generalized coordinate and conjugate momentum. Of course, this degree of freedom is
absent in polarized $U(1)$ models. However the algebraic coordinate conditions which 
we have imposed in fixing the
metric form (1) (essentially zero shift and a ``harmonic'' time function when the
choice $N = e^\Lambda$ is enforced) are not completely rigid since they depend
still upon the choice of an initial hypersurface. There are many other harmonic
time functions (i.e., solutions of the wave equation having timelike gradient)
which could have been used instead of any given one without disturbing the
metric form but one can analyze these asymptotically using the formal expansion
methods of GM. This (not yet rigorously justified) analysis determines the
asymptotic behavior of any such time function relative to a given one (with
respect to which AVTD behavior of the metric is assumed) and shows that
asymptotic velocity term dominance would be preserved in any harmonic time,
zero shift gauge provided it holds in the original one. Thus the occurence of
AVTD behavior does not seem to depend upon the choice of a preferred initial
hypersurface, at least not within the class of gauges under study.

\section*{Acknowledgements}
We would like to thank the Albert Einstein Institute at Potsdam for
hospitality. This work was supported in part by National Science Foundation
Grants PHY9507313 and PHY9503133. Numerical simulations were performed at the
National Center for Supercomputing Applications (University of Illinois).

\vfill
\eject

\section*{Figure Captions}
\bigskip
Figure 1. Convergence of the constraints. The Hamiltonian (triangles), $u$-momentum
(circles), and $v$-momentum constraints vs $\tau$ for spatial resolutions of $128
\times 128$ (broken line),
$248 \times 248$ (solid line), and $512 \times 512$ (dashed line). (a) A second order (in
time) accurate symplectic PDE solver yields a decrease in the magnitude of the
constraints by a factor of $4$ as the number of spatial grid points in each
direction is doubled (second order convergence). (b) A fourth order (in time) accurate
symplectic PDE solver
\cite{suzuki90,suzuki91} yields a factor of $16$ decrease in the magnitude of the
constraints (fourth order convergence). Only the early part of the evolution is shown.
Note that in addition to improvement in convergence with increasing spatial resolution,
one also finds convergence with increasing time order accuracy. In all cases, the spatial
differencing scheme \cite{norton92} is fourth order accurate and centered about the
spatial grid point of interest \cite{choptuik91}.

\bigskip

Figure 2. Exponential decay of the spatial derivative terms in the Hamiltonian.
$\log_{10} |V|$ vs $\tau$ is displayed for two spatial points (solid line). The
straight line slope indicates exponential decay. In each case, the corresponding
values of $(-p_\Lambda + p_z/2) \tau /\ln10$ are also displayed (dashed line). Figures
2--5 are obtained from a second order accurate PDE solver with $512^2$ spatial grid
points.

\bigskip

Figure 3. Decay of the deviations from constant in time behavior of quantities
constant in the AVTD limit. The maximum value of the change with time over the
spatial grid of the quantities $p_x$, $c_z$, $v_z$, $c_\varphi$, $v_\varphi$, and
$p_\Lambda$ are shown vs $\tau$. The non-vanishing values at late times are at the
level of machine precision errors.

\bigskip

Figure 4. Comparison of predicted and measured values of the variables (a)
$\Lambda$, (b) $z$, (c) $\varphi$, and (d) $x$ vs $\tau$ are shown for two
representative spatial points. For (a), (b), and (c), the data is represented by circles
or squares while the solid line is a linear fit to the data. For (d), the fit is to $x =
a + b e^{-p_z \tau}$ as is consistent with (\ref{avtdsoln}).

\begin{figure}[bth]
\begin{center}
\makebox[4in]{\psfig{file=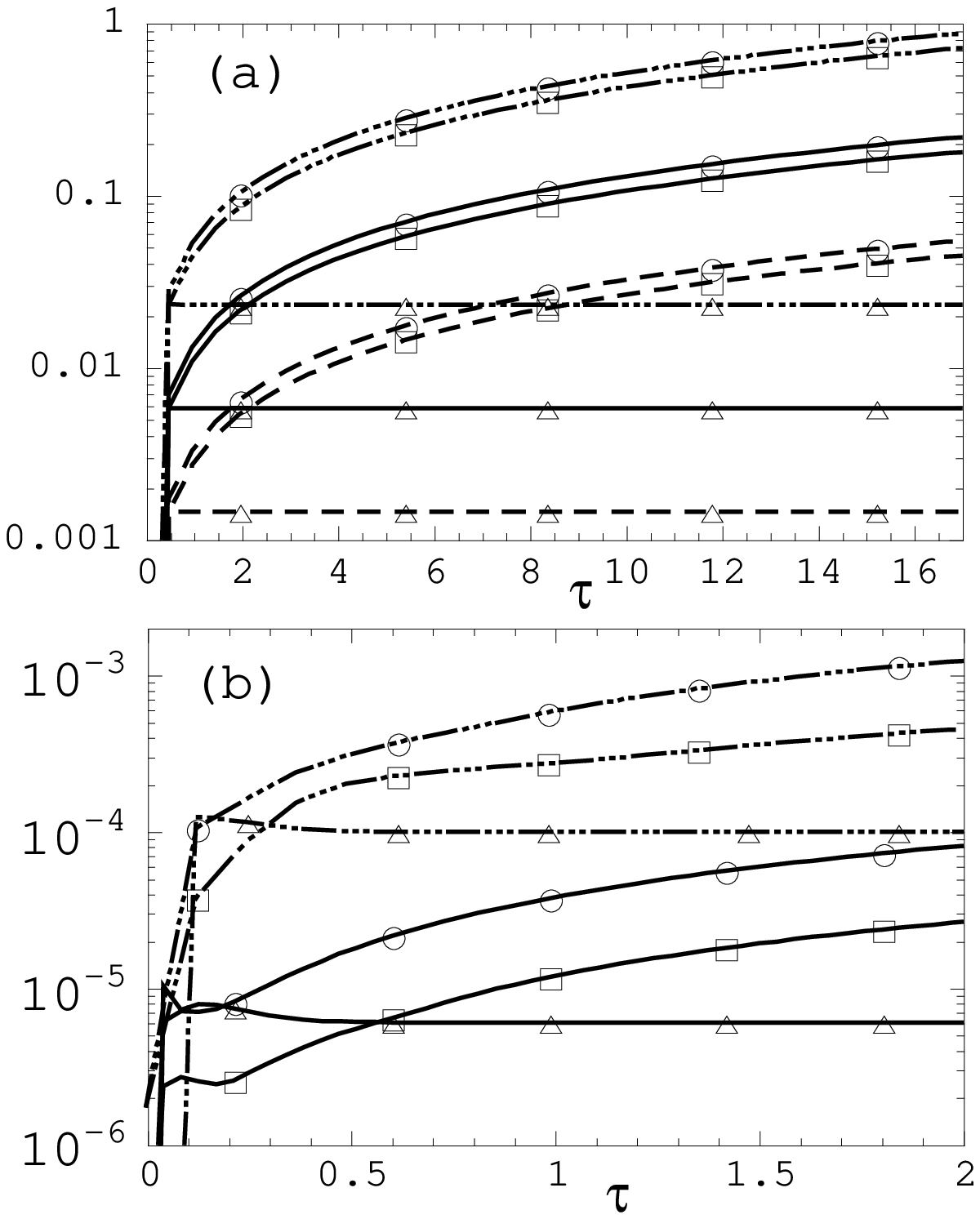,width=3.5in}}
\caption{}
\end{center}
\end{figure}

\begin{figure}[bth]
\begin{center}
\makebox[4in]{\psfig{file=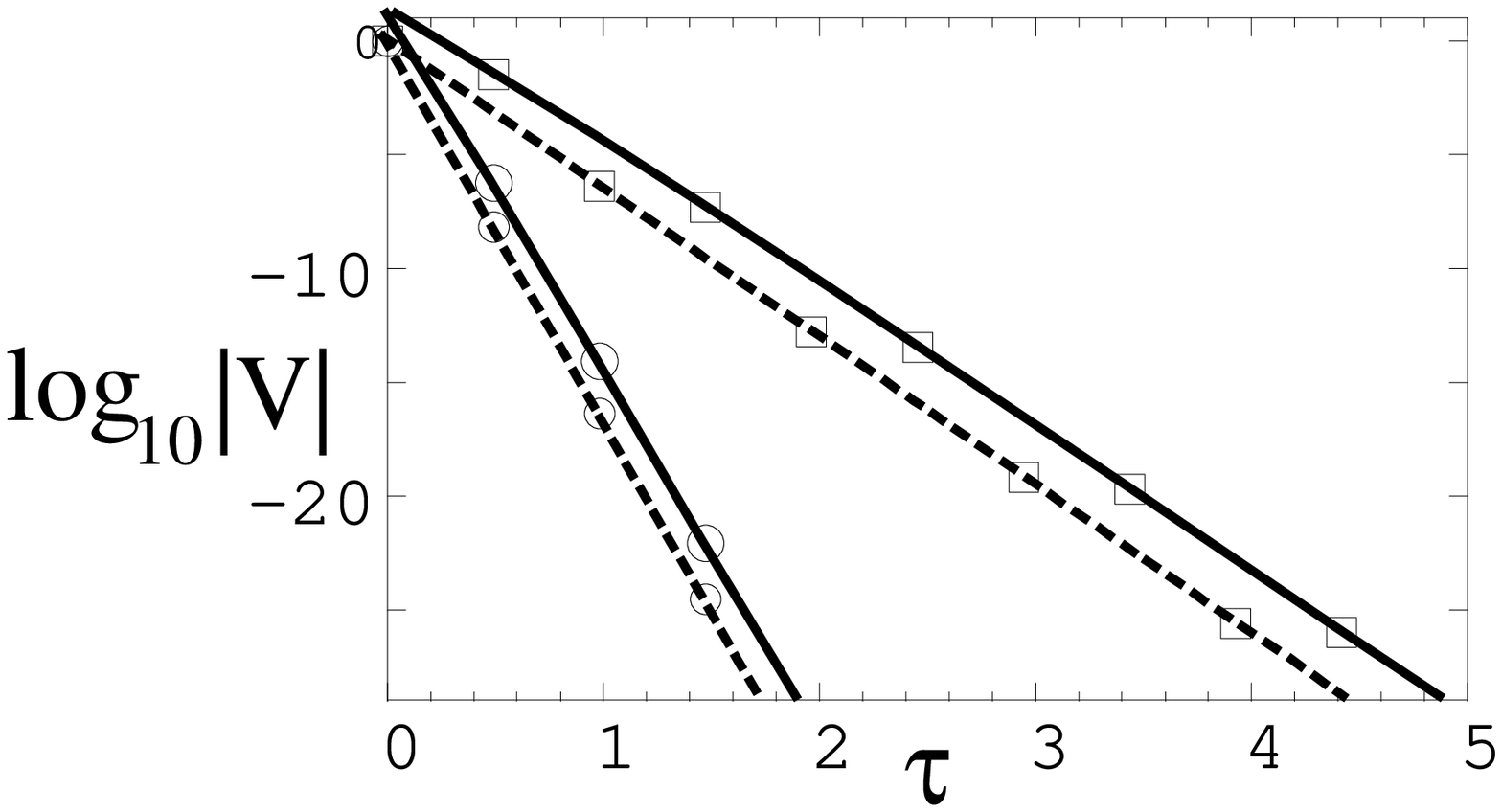,width=3.5in}}
\caption{}
\end{center}
\end{figure}

\begin{figure}[bth]
\begin{center}
\makebox[4in]{\psfig{file=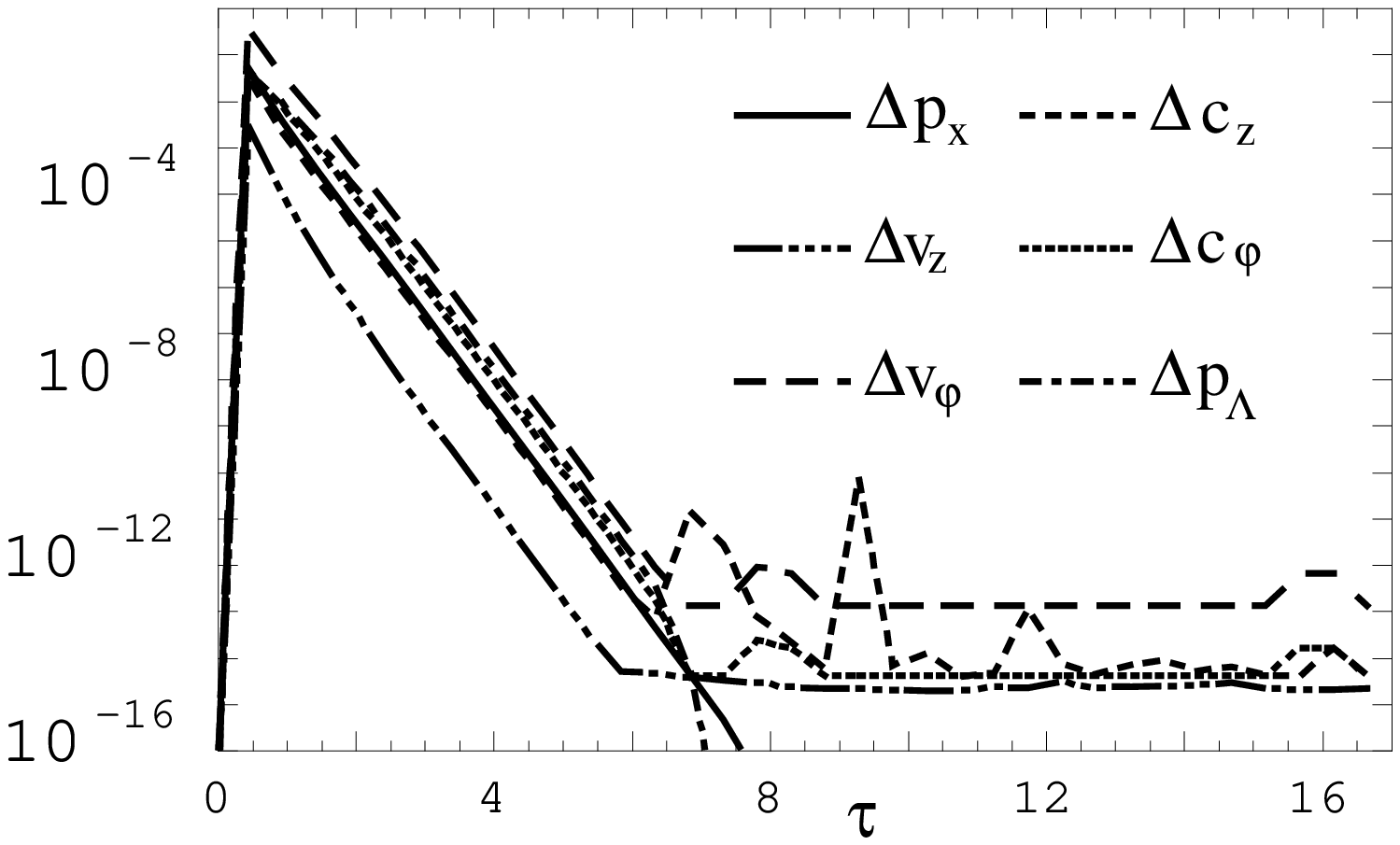,width=3.5in}}
\caption{}
\end{center}
\end{figure}

\begin{figure}[bth]
\begin{center}
\makebox[4in]{\psfig{file=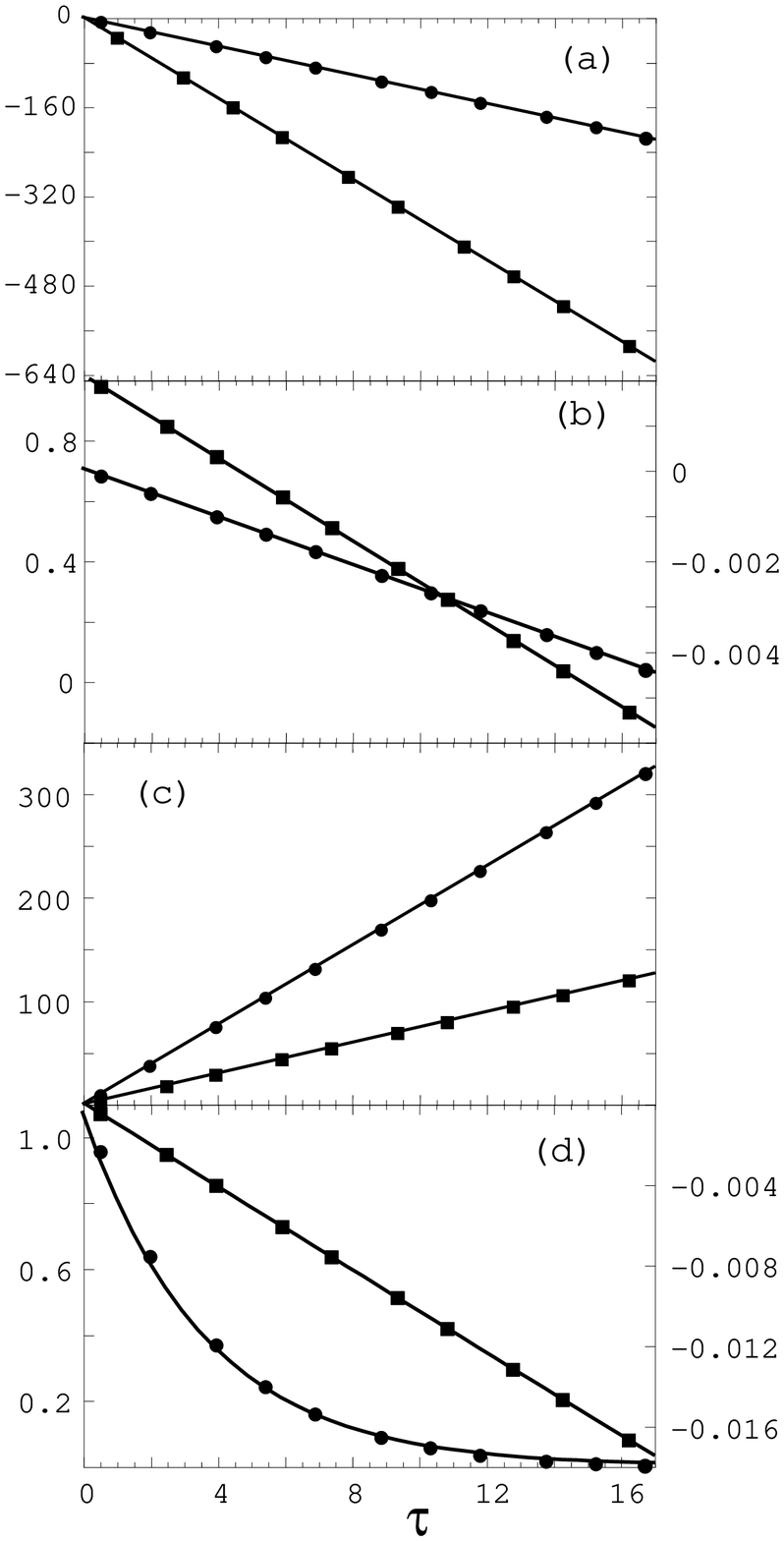,width=3.5in}}
\caption{}
\end{center}
\end{figure}

\end{document}